\PassOptionsToPackage{unicode}{hyperref}
\PassOptionsToPackage{hyphens}{url}
\PassOptionsToPackage{dvipsnames,svgnames,x11names}{xcolor}
\documentclass[
  11pt,
]{article}
\usepackage{xcolor}
\usepackage[margin=1in]{geometry}
\usepackage{amsmath,amssymb}
\usepackage{iftex}
\ifPDFTeX
  \usepackage[T1]{fontenc}
  \usepackage[utf8]{inputenc}
  \usepackage{textcomp} % provide euro and other symbols
\else % if luatex or xetex
  \usepackage{unicode-math} % this also loads fontspec
  \defaultfontfeatures{Scale=MatchLowercase}
  \defaultfontfeatures[\rmfamily]{Ligatures=TeX,Scale=1}
\fi
\usepackage{lmodern}
\ifPDFTeX\else
\fi
\IfFileExists{upquote.sty}{\usepackage{upquote}}{}
\IfFileExists{microtype.sty}{% use microtype if available
  \usepackage[]{microtype}
  \UseMicrotypeSet[protrusion]{basicmath} % disable protrusion for tt fonts
}{}
\makeatletter
\@ifundefined{KOMAClassName}{% if non-KOMA class
  \IfFileExists{parskip.sty}{%
    \usepackage{parskip}
  }{% else
    \setlength{\parindent}{0pt}
    \setlength{\parskip}{6pt plus 2pt minus 1pt}}
}{% if KOMA class
  \KOMAoptions{parskip=half}}
\makeatother
\usepackage{longtable,booktabs,array}
\usepackage{caption}
\usepackage{calc} % for calculating minipage widths
\usepackage{etoolbox}
\makeatletter
\patchcmd\longtable{\par}{\if@noskipsec\mbox{}\fi\par}{}{}
\makeatother
\usepackage{graphicx}
\makeatletter
\newsavebox\pandoc@box
\newcommand*\pandocbounded[1]{% scales image to fit in text height/width
  \sbox\pandoc@box{#1}%
  \Gscale@div\@tempa{\textheight}{\dimexpr\ht\pandoc@box+\dp\pandoc@box\relax}%
  \Gscale@div\@tempb{\linewidth}{\wd\pandoc@box}%
  \ifdim\@tempb\p@<\@tempa\p@\let\@tempa\@tempb\fi% select the smaller of both
  \ifdim\@tempa\p@<\p@\scalebox{\@tempa}{\usebox\pandoc@box}%
  \else\usebox{\pandoc@box}%
  \fi%
}
\def\fps@figure{htbp}
\makeatother
\providecommand{\tightlist}{%
  \setlength{\itemsep}{0pt}\setlength{\parskip}{0pt}}
\usepackage{xurl}
\usepackage{bookmark}
\IfFileExists{xurl.sty}{\usepackage{xurl}}{} % add URL line breaks if available
\makeatletter
\@ifundefined{xmpquote}{}{}
\makeatother
\hypersetup{
  pdftitle={TEA-Bench-BR: Can Coding Agents Turn a Branching Clinical Specification into Correct Code?},
  pdfauthor={Marcio Andre de Souza Estevam (ORCID 0009-0004-8551-3231)},
  colorlinks=true,
  linkcolor={blue},
  filecolor={Maroon},
  citecolor={Blue},
  urlcolor={blue},
  pdfcreator={LaTeX via pandoc}}

\title{TEA-Bench-BR: Can Coding Agents Turn a Branching Clinical
Specification into Correct Code?}
\usepackage{etoolbox}
\makeatletter
\providecommand{\subtitle}[1]{% add subtitle to \maketitle
  \apptocmd{\@title}{\par {\large #1 \par}}{}{}
}
\makeatother
\subtitle{A synthetic, regenerable and contamination-aware benchmark
built on Brazil's autism screening pathway}
\author{Marcio Andre de Souza Estevam (ORCID
\href{https://orcid.org/0009-0004-8551-3231}{0009-0004-8551-3231})}
\date{}

\begin{document}
\maketitle
\begin{abstract}
Coding agents increasingly implement rules written by domain experts,
and bugs in such code are silent: it runs and returns a plausible
answer. We introduce TEA-Bench-BR, a benchmark built on the Brazilian
pathway for early autism (TEA) screening: three fictitious instruments
with the structure of real ones, a multi-professional referral flow, and
export to FHIR R4 with the national RNDS profiles, all specified in
Portuguese. A generator samples each task's parameters (144 tasks, 6,798
hidden cases), and the hidden cases kill all 1,955 non-equivalent
mutants of the reference. With Gemma 4 as the agent, the 12B solves
38.0\% of single-instrument tasks and none at the FHIR level; the 31B
solves 77.8\%, 85.2\% and 78.7\% at levels 1 to 3 and passes over 99\%
of the hidden cases, its remaining failures mostly in input validation
rather than clinical decisions. With the same tasks in English, the 31B
solves 57.4\% at level 3, as its code validates an identifier less
often; the 12B's, more often. The 12B misses 10.10\% of the cases that
call for referral; the 31B, 0.02\%. A contamination probe on a modified
PHQ-9 finds no error that follows the original rules.
\end{abstract}

\subsection{1. Introduction}\label{introduction}

Benchmarks for coding agents mostly measure whether an agent can fix a
bug in an existing repository, as in SWE-bench {[}1{]} and the main
track of this competition. A different and common job is
\emph{specification-to-code}: a domain expert writes the rules and the
agent implements them. A clinical screening questionnaire has
reverse-scored items, cut-offs, rules for missing answers and a second
stage that only runs for some children. Each rule is simple; together
they form a branching program whose mistakes do not crash:
\texttt{\textgreater{}} instead of \texttt{≥} at a threshold silently
classifies a child at risk as low risk.

In Brazil, autism screening in primary care involves several
professions, and results flow into the national health data network
(RNDS) as FHIR R4 resources. An agent helping a municipal health team
must get scoring, referral rules and the interoperability format right
at once, from a Portuguese specification.

Evaluating this is hard for three reasons: real instruments such as the
M-CHAT-R/F {[}2{]} are copyrighted and cannot be redistributed; popular
ones are all over the web, so a model may answer from memory; and a few
hand-written tests let wrong code pass, as EvalPlus showed {[}3{]}.

TEA-Bench-BR addresses all three. Its contributions are:

\begin{enumerate}
\def\labelenumi{\arabic{enumi}.}
\tightlist
\item
  \textbf{A synthetic but structurally faithful benchmark} of three
  instruments, a referral flow and a FHIR export, at three levels of
  conditional complexity, plus a contamination probe built on the
  public-domain PHQ-9 {[}4{]}.
\item
  \textbf{A parameterized generator.} Every task samples its own
  cut-offs, reverse items, age ranges and follow-up rules, so the
  benchmark can be regenerated whenever it leaks; we regenerate it once
  and measure that it stays valid.
\item
  \textbf{Mutation-validated hidden tests.} For every task, the hidden
  cases kill each mutant carrying one typical implementation error,
  while visible tests alone would accept most wrong attempts.
\item
  \textbf{An evaluation of Gemma 4} 12B and 31B, with analyses of
  clinically unsafe errors, of the specification's language and of
  memory-driven errors.
\end{enumerate}

Regenerable tasks graded by hidden cases can also serve as verifiable
rewards for training agents.

\subsection{2. Related work}\label{related-work}

\textbf{Agent and code benchmarks.} SWE-bench {[}1{]} evaluates
repository-level bug fixing. LiveCodeBench {[}5{]} limits contamination
by collecting problems published after a model's training cut-off.
EvalPlus {[}3{]} showed that 80× more tests lowered HumanEval pass@k by
up to 28.9\%: test adequacy is part of benchmark validity. We report
mutation analysis, a classical measure of test adequacy {[}6{]}, for
every task.

\textbf{Clinical computation.} MedCalc-Bench {[}7{]} has LLMs compute 55
medical scores from patient notes; we evaluate the \emph{code} an agent
writes from a specification. FHIR-AgentBench {[}8{]} has agents answer
questions over FHIR records; we evaluate the opposite direction,
producing FHIR resources a validator accepts.

\textbf{Contamination and templated benchmarks.} GSM-Symbolic {[}9{]}
finds that accuracy drops when only the numbers of templated GSM8K
problems change, a sign of pattern matching. Our generator applies the
same idea to specifications, and our contamination probe goes further:
it keeps a famous instrument's name but changes its rules.

\subsection{3. Benchmark design}\label{benchmark-design}

\pandocbounded{\includegraphics[keepaspectratio,alt={Figure
1}]{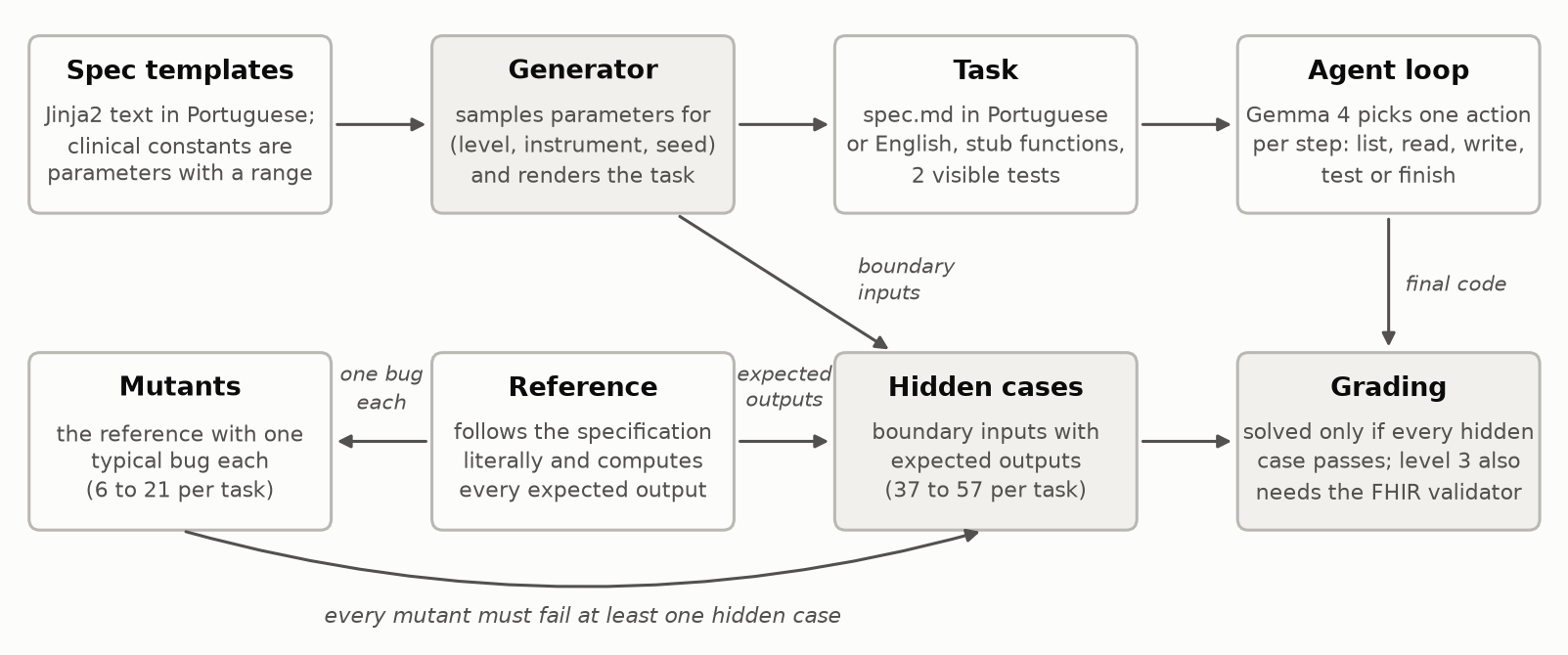}}

\textbf{Figure 1.} How a task is built and how an attempt is graded.

\subsubsection{3.1 Instruments and levels}\label{instruments-and-levels}

All instruments are fictitious, written in Portuguese by the author and
reviewed by a psychologist and a speech therapist for plausibility and
for distance from protected instruments: the review moved the RICS-20
defaults away from the scoring structure of the M-CHAT-R/F, rewrote IPSQ
wording that echoed the Sensory Profile, and fixed three clinically
wrong follow-up questions. A shared conventions page fixes age in months
(integer part of days ÷ 30.4375), half-up rounding, strict answer types
(\texttt{True}, \texttt{1.0} and \texttt{"3"} are not the integer 3),
and that clinical conditions (age out of range, too many missing
answers) return a status, never an exception.

\begin{itemize}
\tightlist
\item
  \textbf{RICS-20} (social communication): 20 yes/no items, some
  reverse-worded; missing answers count as failures up to a limit; risk
  bands; a follow-up interview for each failed item, with
  typical-example and red-flag questions.
\item
  \textbf{IPSQ} (sensory processing): Likert items in four interleaved
  quadrants, reverse-scored items, imputation of missing answers by the
  rounded quadrant mean, and per-quadrant bands.
\item
  \textbf{EML} (language milestones): expected ages, key milestones, a
  regression question that triggers immediate referral, and age
  correction for preterm children.
\item
  \textbf{Referral flow}: checks that each result was signed by an
  allowed professional with a matching council and that the facility
  code is valid, then combines results into referrals (including
  re-screening at the earlier of two deadlines) and a priority.
\item
  \textbf{FHIR export}: a \texttt{Bundle} with a \texttt{Patient}
  conforming to the RNDS \texttt{BRIndividuo-1.0} profile, one
  \texttt{Observation} per usable instrument and one
  \texttt{ServiceRequest} per referral.
\end{itemize}

Levels increase in conditional depth (Table 1); level 4 is the
contamination probe.

\textbf{Table 1.} Benchmark composition.

{\def\LTcaptype{none} % do not increment counter
\begin{longtable}[]{@{}
  >{\raggedright\arraybackslash}p{(\linewidth - 6\tabcolsep) * \real{0.2500}}
  >{\raggedright\arraybackslash}p{(\linewidth - 6\tabcolsep) * \real{0.2500}}
  >{\raggedright\arraybackslash}p{(\linewidth - 6\tabcolsep) * \real{0.2500}}
  >{\raggedright\arraybackslash}p{(\linewidth - 6\tabcolsep) * \real{0.2500}}@{}}
\toprule\noalign{}
\begin{minipage}[b]{\linewidth}\raggedright
Level
\end{minipage} & \begin{minipage}[b]{\linewidth}\raggedright
Agent implements
\end{minipage} & \begin{minipage}[b]{\linewidth}\raggedright
Tasks
\end{minipage} & \begin{minipage}[b]{\linewidth}\raggedright
Hidden cases
\end{minipage} \\
\midrule\noalign{}
\endhead
\bottomrule\noalign{}
\endlastfoot
1 & One instrument's scoring (RICS-20 stage 1, IPSQ or EML) & 36 &
1,680 \\
2 & RICS-20 with the follow-up stage & 36 & 1,734 \\
3 & Referral flow + FHIR export & 36 & 2,052 \\
4 & PHQ-9 real / synthetic / trap (12 triplets) & 36 & 1,332 \\
\end{longtable}
}

\subsubsection{3.2 Parameterized
generation}\label{parameterized-generation}

Each instrument is a Jinja2 template whose clinical constants (cut-offs,
reverse items) are parameters with an allowed range. A task is a (level,
instrument, seed) triple: the generator samples valid parameters,
renders the specification and builds two independent input batches, one
for the hidden cases and one for five worked examples and two visible
tests, with no input in common.

Expected outputs come from a reference implementation that follows the
specification literally, never from hand. Inputs hit every boundary:
first and last day of each age limit, scores at and one below each
cut-off, missing answers at the tolerated limit, invalid values
(\texttt{"SIM"}, \texttt{True}), a mean ending in exactly .5, and
malformed dates.

\subsubsection{3.3 Grading}\label{grading}

A task is solved only if \emph{all} hidden cases pass. Outputs are
compared field by field, with booleans not equal to integers. A FHIR
Bundle passes only if (i) a canonical summary (patient, observations,
referrals; ignoring ids and order) matches the reference, and (ii) the
official HL7 validator, loaded with the RNDS package, reports no
structural error. The RNDS package (BRCore 1.0.0) has two terminology
defects that make even a correct \texttt{Patient} fail; we discount
exactly those two messages through a fixed, published list.

\subsubsection{3.4 Test adequacy}\label{test-adequacy}

For each task we generate 6 to 21 mutants of the reference with one
typical error each, from hand-written operators
(\texttt{generator/mutantes.py}): a \texttt{\textgreater{}} for
\texttt{≥} at a cut-off, 30-day months, banker's rounding in imputation,
forgotten reverse items, missing answers treated as typical, an
unchecked professional council, the wrong FHIR priority code, and, for
level 4, the original PHQ-9 cut-offs. Mutants that cannot behave
differently under a task's parameters are excluded by a condition on the
parameters (for example, banker's rounding when no .5 mean is possible).
The hidden cases kill \textbf{1,955 of 1,955} non-equivalent mutants.
One level-3 mutant (a \texttt{Patient} without \texttt{identifier.use})
is killed only by the validator, which justifies the structural check.
Since operators and cases target the same boundaries, this shows that
the listed error classes are covered, not that every bug is.

\textbf{Regeneration.} We regenerated all 144 tasks with new seeds: the
reference solves every one through the agent path, and the new hidden
cases kill 1,968 of 1,969 mutants. The survivor (30-day months in one
re-screening date) is why adequacy is measured for each regeneration,
not assumed.

\subsection{4. Experiments}\label{experiments}

\textbf{Agent.} Gemma 4 in the quantized variants allowed by the main
competition, \texttt{12B} and \texttt{31B} (\texttt{-it-qat-w4a16-ct}),
served by vLLM 0.33.0. Portuguese runs used Kaggle L4×4 GPUs; English
runs, and a Portuguese control for the 12B, used a rented A100 with the
same settings and the public Hugging Face weights. The loop is minimal
and identical for all runs: at each step the model emits one JSON action
(\texttt{list}, \texttt{read}, \texttt{write} under \texttt{src/},
\texttt{test}, \texttt{finish}), with at most 30 steps, 200k tokens per
attempt and 4,096 tokens per response. Format errors are returned with
what to fix, and a third identical response ends the attempt. Each
attempt runs in a temporary copy of the task. We run 3 attempts per task
with the sampling recommended for Gemma 4 (temperature 1.0, top-p 0.95,
top-k 64), no fixed seed, thinking disabled and vLLM's Gemma 4 reasoning
parser.

\textbf{E1, complexity.} Solve rate by level and model size. As a
secondary rate we also report tasks solved when the six strict-typing
cases (such as a boolean given as an answer, or a facility code given as
a number) are set aside.

\textbf{E2, language.} Each level 1--3 task has an English twin with the
same parameters, examples, visible tests and hidden cases. Only the
prose changes; identifiers (function names, keys and output codes such
as \texttt{"sim"}), visible-test case names and the agent's system
prompt stay in Portuguese, so the English condition is mixed. We compare
languages attempt by attempt with an exact sign test, and ran the 31B's
English level-3 tasks twice.

\textbf{E3, clinical safety.} Among failed hidden cases, a \emph{false
negative} is a case where the reference indicates risk or referral
(medium/high risk, follow-up positive, referral to speech or
occupational therapy, referral for diagnostic evaluation) and the
agent's code indicates none. We report false negatives as a fraction of
positive cases and false positives as a fraction of negative cases.

\textbf{E4, contamination.} Each triplet shares the PHQ-9 structure.
\emph{Real} uses the PHQ-9 cut-offs and item order. \emph{Synthetic}
uses an invented name, four altered cut-offs and shuffled items.
\emph{Trap} keeps the PHQ-9 name and item order but states the
\emph{same} altered cut-offs as the synthetic version. For each failed
synthetic or trap case we check whether the output equals what the
original PHQ-9 rules give, an error that ``follows memory''.

\subsection{5. Results}\label{results}

\pandocbounded{\includegraphics[keepaspectratio,alt={Figure
2}]{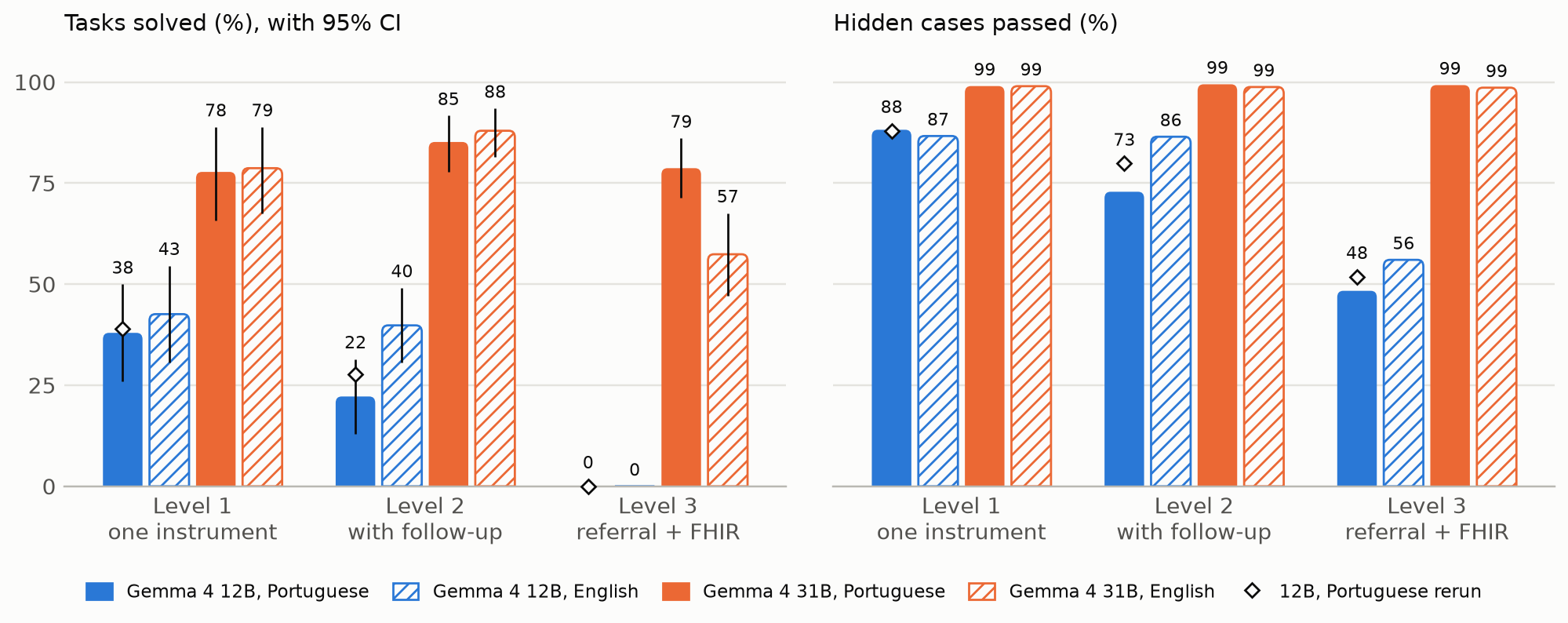}}

\textbf{Figure 2.} Tasks solved and hidden cases passed, by level, model
and specification language. Diamonds: the 12B's Portuguese tasks rerun
on the English runs' machine.

\textbf{Table 2.} Levels 1 to 3 in Portuguese (\%). Solve rates are
means over 3 attempts, with 95\% bootstrap intervals over tasks.

{\def\LTcaptype{none} % do not increment counter
\begin{longtable}[]{@{}
  >{\raggedright\arraybackslash}p{(\linewidth - 8\tabcolsep) * \real{0.2000}}
  >{\raggedright\arraybackslash}p{(\linewidth - 8\tabcolsep) * \real{0.2000}}
  >{\raggedright\arraybackslash}p{(\linewidth - 8\tabcolsep) * \real{0.2000}}
  >{\raggedright\arraybackslash}p{(\linewidth - 8\tabcolsep) * \real{0.2000}}
  >{\raggedright\arraybackslash}p{(\linewidth - 8\tabcolsep) * \real{0.2000}}@{}}
\toprule\noalign{}
\begin{minipage}[b]{\linewidth}\raggedright
Model
\end{minipage} & \begin{minipage}[b]{\linewidth}\raggedright
Measure
\end{minipage} & \begin{minipage}[b]{\linewidth}\raggedright
Level 1
\end{minipage} & \begin{minipage}[b]{\linewidth}\raggedright
Level 2
\end{minipage} & \begin{minipage}[b]{\linewidth}\raggedright
Level 3
\end{minipage} \\
\midrule\noalign{}
\endhead
\bottomrule\noalign{}
\endlastfoot
12B & Tasks solved & 38.0 (25.9--50.0) & 22.2 (13.0--31.5) & 0.0 \\
12B & Solved, typing cases set aside & 52.8 & 22.2 & 0.9 \\
12B & Hidden cases passed & 88.2 & 72.9 & 48.4 \\
31B & Tasks solved & 77.8 (65.7--88.9) & 85.2 (77.8--91.7) & 78.7
(71.3--86.1) \\
31B & Solved, typing cases set aside & 93.5 & 85.2 & 82.4 \\
31B & Solved in ≥ 1 / all 3 attempts & 88.9 / 69.4 & 100.0 / 63.9 &
100.0 / 47.2 \\
31B & Hidden cases passed & 99.0 & 99.5 & 99.3 \\
\end{longtable}
}

\textbf{Complexity (E1).} The 12B degrades with conditional depth and
solves no level-3 task. It ends 53 of 108 level-3 attempts on the token
budget and 17 on repetition, with wrong decisions (76 attempts) and code
that does not run (an undefined name in 19, an unimplemented function in
13). The 31B shows no such drop. It barely uses the loop: 4.2 steps, 1.1
writes and 1.1 test runs per attempt at every level, always ending by
its own \texttt{finish}. Its level-1 average hides a split: it solves
every EML and RICS-20 task but 33.3\% of IPSQ tasks, mostly because it
accepts the boolean \texttt{True} as an answer (23 of 108 level-1
attempts), which the conventions exclude. A \texttt{True} silently read
as 1 is a real defect in clinical data, so the official rate keeps these
cases; set aside, the IPSQ rate is 80.6\%. At level 3 all of its
failures are in the decision function, none in the FHIR export, mostly a
facility code (CNES) given with a letter (10 of 108 attempts) or as a
number (6). The 31B solved every level-2 and level-3 task in at least
one attempt, so the specifications suffice, but only 63.9\% and 47.2\%
in all three: the hidden cases separate reliable from occasional
success. Visible tests do not: every 31B attempt ended with them
passing, yet 19.4\% fail a hidden case, and so do 67.2\% of the 198 12B
attempts that ended green.

\textbf{Language (E2).} The effect of the specification's language
depends on the model (Figure 2). The 31B solves as many level-1 and
level-2 tasks in English (78.7\% and 88.0\%) but fewer at level 3:
57.4\% against 78.7\% (38 attempts solved only in Portuguese against 15
only in English, sign test p = 0.002). An earlier run of the same
English level-3 tasks, on Kaggle, gave 50.0\%. Both languages state that
the facility code (CNES) is a text of 7 digits, yet the final code
checks the digits in 98 of 108 Portuguese attempts and in 70 of 108
English ones in each of the two runs (p \textless{} 0.001); every
attempt without the check fails the case with a letter. The 12B moves
the other way. Rerun in Portuguese on the English runs' machine, it
checks the CNES digits in 50 attempts, against 87 in English (p
\textless{} 0.001), and solves 27.8\% of level-2 tasks (22.2\% on
Kaggle), against 39.8\% in English (17 against 30, p = 0.08).

\textbf{Safety (E3).} The 31B produced 1 false negative (0.02\% of
positive cases) and 1 false positive (0.02\% of negative cases). The 12B
produced 630 false negatives (10.10\%) and 115 false positives (1.79\%),
but the asymmetry is concentrated at level 2, where it misses 25.20\% of
the positive cases; at level 3 it produces more false positives than
false negatives (2.86\% against 0.86\%).

\textbf{Contamination (E4).} Official solve rates for real, synthetic
and trap tasks were 47.2\%, 16.7\% and 52.8\% for the 31B and 11.1\%,
8.3\% and 27.8\% for the 12B, but the differences come from the
strict-typing cases: with them set aside, the 31B solves 100.0\% of
every variant and the 12B 88.9\%, 88.9\% and 83.3\%. None of the errors
on synthetic or trap tasks follows the original PHQ-9 rules (0 of 48
comparable errors for the 31B, 0 of 33 for the 12B).

\subsection{6. Limitations and ethics}\label{limitations-and-ethics}

The instruments are fictitious and not clinically validated; the
benchmark measures fidelity to a specification, not clinical validity,
and must not be used to screen children. No real patient data is used;
all identifiers are synthetic and marked. Protected instruments are not
reproduced: the clinical review checked the Portuguese item wording,
order, examples and default scoring structure, and the English
translation was screened against the same instruments and reworded where
it came close. The PHQ-9 appears only through its public-domain scoring
rules and item domains. The harness was not calibrated on the main
competition's bug-fixing set, which its whole-file protocol does not
support; the 31B passing over 99\% of the hidden cases through it argues
against a harness artifact. Three rules were made explicit after earlier
runs showed the 31B missing them: strict answer types, the order of
checks for an unknown item or milestone, and that a field with only
spaces counts as empty. The hidden cases already checked them and did
not change; only runs with the final specification are reported.
Sampling had no fixed seed, so attempts are not reproducible bit for
bit. Attempts whose code references the hidden tests or the reference
implementation are flagged and excluded (none were). The benchmark was
written with the help of an AI coding assistant (Claude Code) and
reviewed by the author; the evaluated agent is Gemma 4 only.

\textbf{Availability.} Specifications, generator, reference, harness and
results:
\url{https://www.kaggle.com/datasets/marcioestevampsi/tea-bench-br-clinical-spec-to-code-benchmark}
(code MIT, data CC-BY-4.0).

\subsection{References}\label{references}

{[}1{]} C. E. Jimenez et al.~SWE-bench: Can Language Models Resolve
Real-World GitHub Issues? ICLR 2024. arXiv:2310.06770.\\
{[}2{]} D. L. Robins et al.~Validation of the Modified Checklist for
Autism in Toddlers, Revised with Follow-up (M-CHAT-R/F).
\emph{Pediatrics} 133(1):37--45, 2014.\\
{[}3{]} J. Liu et al.~Is Your Code Generated by ChatGPT Really Correct?
Rigorous Evaluation of Large Language Models for Code Generation.
NeurIPS 2023. arXiv:2305.01210.\\
{[}4{]} K. Kroenke et al.~The PHQ-9: Validity of a Brief Depression
Severity Measure. \emph{J Gen Intern Med} 16(9):606--613, 2001.\\
{[}5{]} N. Jain et al.~LiveCodeBench: Holistic and Contamination Free
Evaluation of Large Language Models for Code. arXiv:2403.07974, 2024.\\
{[}6{]} R. A. DeMillo et al.~Hints on Test Data Selection: Help for the
Practicing Programmer. \emph{Computer} 11(4):34--41, 1978.\\
{[}7{]} N. Khandekar et al.~MedCalc-Bench: Evaluating Large Language
Models for Medical Calculations. NeurIPS 2024 Datasets and Benchmarks.
arXiv:2406.12036.\\
{[}8{]} G. Lee et al.~FHIR-AgentBench: Benchmarking LLM Agents for
Realistic Interoperable EHR Question Answering. ML4H 2025.
arXiv:2509.19319.\\
{[}9{]} I. Mirzadeh et al.~GSM-Symbolic: Understanding the Limitations
of Mathematical Reasoning in Large Language Models. arXiv:2410.05229,
2024.\\

\end{document}